\documentclass[11pt]{article}

\usepackage[utf8]{inputenc}
\usepackage{amsmath,amssymb}
\usepackage{graphicx}
\usepackage{hyperref}
\usepackage{authblk}
\usepackage[margin=1in]{geometry}
\usepackage{caption}
\usepackage{subcaption}

\title{The Study of Human Preference Based on Integrated Analysis of N1 and LPP Components}

\author[1]{Siyuan Li}
\author[1]{Xiangze Meng}
\author[1]{Yijian Yang}
\author[1]{Yiwen Xu}
\author[1]{Yunfei Wang}
\author[1]{Chenghu Qiu}
\author[1]{Hanyi Jiang}
\author[1]{Pin Wu}
\author[1]{Shegnbo Chen}
\author[1]{Xiao Wei}
\author[1]{Hao Wang}
\author[2]{Lan Ni}
\author[1]{Huiran Zhang\thanks{Corresponding author, E-mail address: hrzhangsh@shu.edu.cn}}

\affil[1]{School of Computer Engineering and Science, Shanghai University, 99 Shangda Rd., Baoshan District, Shanghai, 200444, China}
\affil[2]{College of Liberal Arts, Shanghai University, 99 Shangda Rd., Baoshan District, Shanghai, 200444, China}

\date{}

\begin{document}

\maketitle

\begin{abstract}
\normalsize 
Human preference research is a significant domain in psychology and psychophysiology, with broad applications in psychiatric evaluation and daily life quality enhancement. This study explores the neural mechanisms of human preference judgments through the analysis of event-related potentials (ERPs), specifically focusing on the early N1 component and the late positive potential (LPP). Using a mixed-image dataset covering items such as hats, fruits, snacks, scarves, drinks, and pets, we elicited a range of emotional responses from participants, while recording their brain activity via EEG. Our work innovatively combines the N1 and LPP components to reveal distinct patterns across different preference levels. The N1 component, particularly in frontal regions, showed increased amplitude for preferred items, indicating heightened early visual attention. Similarly, the LPP component exhibited larger amplitudes for both preferred and non-preferred items, reflecting deeper emotional engagement and cognitive evaluation. In addition, we introduced a relationship model that integrates these ERP components to assess the intensity and direction of preferences, providing a novel method for interpreting EEG data in the context of emotional responses. These findings offer valuable insights into the cognitive and emotional processes underlying human preferences, and present new possibilities for brain-computer interface applications, personalized marketing, and product design.
\end{abstract}


\section{Introduction}
\label{sec1}
The study of human preferences is an important research direction in psychology and psychophysiology, with widespread applications in Psychiatric evaluation and human life. On a medical level, studies of care preferences for cancer survivors enable more accurate selection of appropriate care types, providing strong support for early intervention and subsequent treatment \cite{WOS:000495477900038}. In daily life, research on human preferences not only contributes to improving individual quality of life \cite{WOS:000089033000018}, but also has great application value in fields such as commerce \cite{9623092}, education, and health \cite{WOS:001087962200001}. As a result, various disciplines, from economics \cite{WOS:000621632000018} to psychology \cite{GOTO201711}, have been exploring the complex factors that shape these preferences. In neuroscience, researchers attempt to reveal the internal mechanisms behind decision-making by analyzing how the brain responds to different product stimuli. In the new research field of BCI (Brain-computer interface), detecting and decoding preferences is an important approach for achieving human-computer interaction and control. By identifying user preference signals, BCI systems can more accurately respond to user needs, thereby improving operational sensitivity and precision \cite{WOS:001124163900031}.

Currently, an increasing number of biometric methods are being used in preference research. For instance, in personalized video recommendation research, functional near-infrared spectroscopy (fNIRS) has been used to measure and judge individual preferences for videos \cite{8229257}. Using functional magnetic resonance imaging (fMRI), Calvert et al. found that it could monitor brain activity patterns to predict consumer reactions to new products, advertisements, etc., offering more accurate forecasts for the marketing industry \cite{6210549}. Due to the high temporal resolution of Electroencephalography (EEG), it is considered more suitable for monitoring real-time changes in preferences and applied in studies of individual preferences increasingly. Research has shown that different product categories may elicit different activity patterns in the brain.  When product preferences are similar, the N200 component reflects preference only in specific contexts where similar products are involved, and fails to reflect consumer preferences in single-reference evaluations \cite{Wang2021HowDR}. These studies typically categorize products into preferred and non-preferred groups and assign corresponding probability scores to indicate the differences between them \cite{Guo2018AffectivePM}. Two experimental paradigms are commonly used in such research: one presents a single product image at a time for participants to evaluate or choose, while the other presents two product images simultaneously for comparison \cite{Wang2021HowDR}. Guo et al. used the former paradigm to analyze user preferences for humanoid robot images through ERP components such as parieto-occipital N1, frontal P2, and LPP, finding that preferred humanoid robot images induced enhanced parieto-occipital N1 and frontal P2 in the early stage and scalp-distributed LPP in the late stage \cite{Guo_Li_Chen_Duffy_2022}. Wang et al. used the latter paradigm to analyze participants’ preferences for different types of fruit and found that under this paradigm, the N200 required specific reference points to reflect participants' preferences \cite{Wang2021HowDR}. These studies reveal unique EEG features related to specific preferences.

Event-Related Potentials (ERPs) provide robust measurements related to cognitive processes and offer objective metrics for evaluating stimulus preferences \cite{Schupp_Junghöfer_Weike_Hamm_2003}. Key EEG frequency bands and ERP components can be used to detect cognitive processes \cite{Wang2021HowDR}. For example, the N2 component is sensitive to negative facial expressions \cite{Rossignol_2002}, and emotional changes are primarily associated with the amplitude of ERP components, with later components reflecting changes in emotional arousal \cite{Olofsson_Nordin_Sequeira_Polich_2008}. Notably, the early posterior negativity (EPN) increases in amplitude in response to stimuli, especially those with high evolutionary significance, while behavioral and electrophysiological responses to task stimuli indicate successful top-down attention control over non-emotional stimuli \cite{Schupp_Junghöfer_Weike_Hamm_2003}. When focusing on just these three ERP components, they can distinguish consumer preferences for specific products to some extent, and the use of single ERP components can infer behavior \cite{Goto_Lim_Shee_Hatano_Khong_Buratto_Watabe_Schaefer_2019}. Late components, representing advanced emotional processes, reflect sustained motivational attention and stimuli requiring attention, including the late positive potential (LPP), late frontal potential (LFP), and slow positive wave (SPW) \cite{Guo_Ding_Wang_Liu_Jin_2016, Guo2018AffectivePM, Istók_Friberg_Huotilainen_Tervaniemi_2013, Langeslag_van_Strien_2018}. Specifically, emotional disturbance scores are negatively correlated with the magnitude of emotion-related ERP activities, inhibiting such activities. This allows ERPs to be used to analyze mental disorders, such as the modulation of early and late emotion-related ERPs in emotional disorder attempts to suppress emotions \cite{Walker_O’Connor_Schaefer_2011}, among others. These studies address consumer preference expression and multidimensional individual experience, as well as the analysis of specific brain regions and ERP components within specific time windows. However, it is necessary to provide a more comprehensive perspective for analyzing multiple ERP components across different brain regions and stimulus types, which is crucial for understanding product preferences.

This research aimed to identify a general component configuration that reflects emotional responses to preferences across different product categories, and to explore the commonalities in human preferences for various items. We developed a mixed-image dataset and had subjects assess their preferences for items in the dataset. The data were analyzed based on varying levels of preference, with a focus on exploring pure preference-inducing ERP component combinations. This study seeks to theoretically analyze the factors influencing preferences independent of stimulus type. In this work, we first selected experimental stimulus samples and designed experiments based on preliminary research. Second, we analyzed brain preference responses based on survey results and experimental data. Finally, we obtained insights into brain activity during preference judgments through analysis and discussion. The findings not only offer new insights into the commonalities of human preferences for different objects, but also demonstrate how the combination of multiple channels and time domains can deeply probe emotional responses in human consciousness, bringing new opportunities for the study of human preferences.

\section{Materials and Methods}
\subsection{Stimulus Samples}

A series of mixed images of real objects was used as experimental stimuli, covering six categories: hats, fruits, snacks, scarves, beverages, and pets. These different types of images were intended to explore people's emotional responses to various items. Specifically, hats represent clothing, reflecting people's self-expression and identity \cite{hats000459829000005}; fruits symbolize health and nature, representing the instinctual pursuit of basic physiological needs like food \cite{foods12112095}; snacks are associated with instant gratification andb   pleasure, activating brain regions related to reward mechanisms \cite{Okpiaifo2023}; scarves serve not only as decorative items but also carry emotional value, showcasing personal style and lifestyle preferences \cite{scarf000787273200001}; beverages reflect the basic human need for hydration, especially drinks like coffee and tea that carry specific emotional and social value, representing preferences for social interaction \cite{foods12112095}; while pets are connected to emotional bonding and companionship \cite{pet_2009}, demonstrating people's need for close relationships and emotional support. These six categories of images were randomly mixed to minimize the potential influence of individual preference biases toward specific item categories.

These items are common in daily life and enjoy a high level of acceptance and recognition across diverse groups of people. As a result, selecting these items as experimental stimuli enhances the generalizability and representativeness of the research findings. Furthermore, these objects are rich in emotional attributes, effectively triggering a range of emotional responses from participants, allowing for a more comprehensive investigation into how different stimuli influence emotions and cognition. To eliminate the potential impact of factors such as image size or background, all images used in the experiment were uniformly adjusted, including image size and background. This standardization ensured consistency in experimental conditions and contributed to the collection of more accurate and reliable data.

\subsection{Experimental Design}

Before the experiment, it was necessary to determine specific details such as rating scales and descriptive terms through survey research. Evaluating whether potential participants could clearly discern their emotional preferences for a product was also crucial for the experiment's outcome. Therefore, a questionnaire survey was designed for this experiment. The survey included abstract questions about the categories of products we selected, such as asking respondents to choose their preferred beverage. These questionnaires were distributed in the form of online surveys to undergraduate students from different academic disciplines to obtain more authentic and generalizable data. Ultimately, we received responses from 43 students, including 34 males and 9 females. Based on the survey results, we decided to use a three-tier emotional scale of "dislike-neutral-preferred" as the emotional grading system for the formal experiment.

\subsubsection{participants}

The formal experiment recruited 14 university students (mean age 21 years, range 20-23 years, SD = 0.76), including 10 males and 4 females, for EEG data collection in the laboratory. Prior to participating in the formal experiment, all participants had completed the survey questionnaire and were therefore familiar with the experimental procedures. During the testing phase, participants were exposed to emotional stimulation tasks similar to those used in the formal experiment, which were designed to assess their ability to distinguish between different emotional categories. The test results showed that all participants were able to accurately identify the emotional categories, demonstrating that they were qualified to participate in the formal experiment. All data collection procedures had received the participants' informed consent, and the experimental process had been publicly announced on the school and local community bulletin boards, where it also received approval. All participants underwent physical and mental health screenings, confirming that they all had normal or corrected-to-normal vision, had no history of neurological or psychiatric disorders, and were all right-handed.

\subsubsection{Experimental environment and equipment}

\begin{figure}[htbp]
    \centering
    \includegraphics[width=0.6\textwidth]{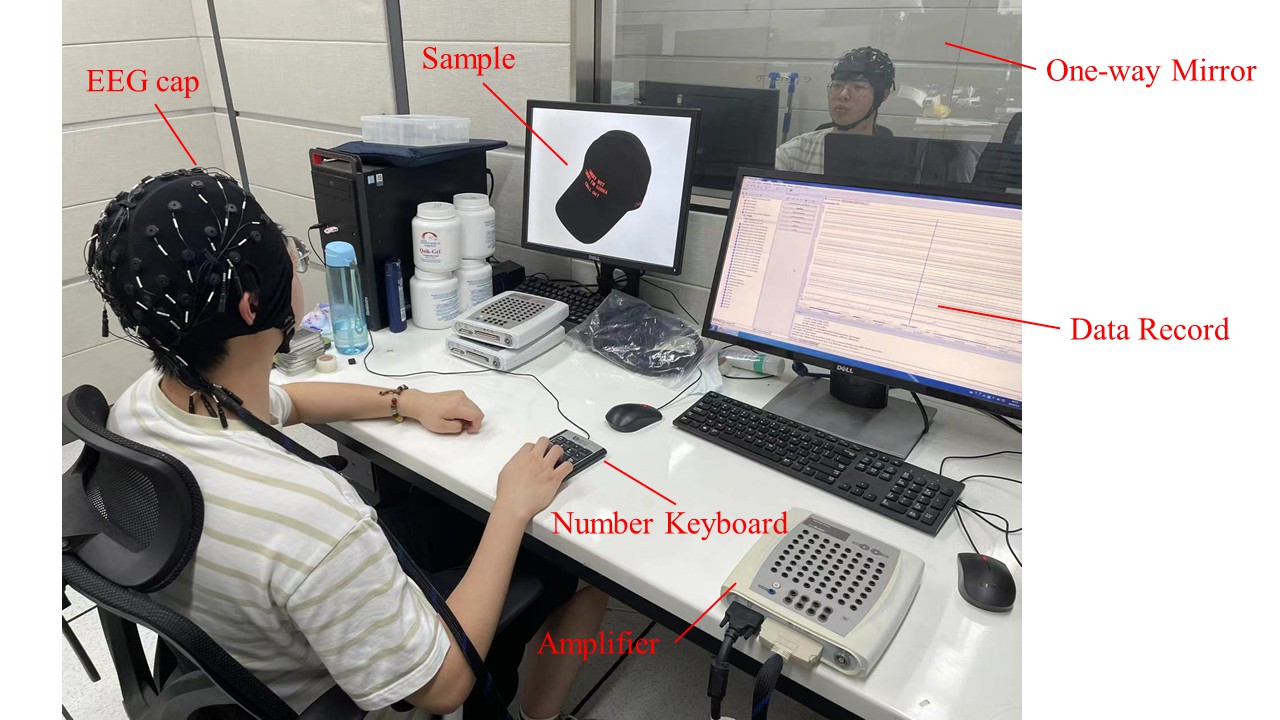}
    \caption{Layout of EEG data acquisition room and equipment}
    \label{fig:env-and-equip}
\end{figure}

The experiment was conducted in the EEG data acquisition room as shown in Fig.  \ref{fig:env-and-equip}. The acquisition room is an isolated chamber with soundproofing and consistent, adequate lighting. To prevent interference from unrelated electromagnetic signals, the use of other electronic devices inside and around the room was strictly prohibited during the experiment. The room is equipped with a one-way glass window, allowing staff to observe the participants from outside without affecting them. Except for the one-way glass, the other walls of the room are opaque.

In this study, EEG data were recorded using a Synamps 2 amplifier (Neuroscan system), a 64-channel electrode cap, and Curry 8.0 software. The electrode cap was placed on the participants' scalps according to the International 10-20 System, ensuring standardized and consistent electrode placement. Fifteen electrodes were selected for recording, including F3, Fz, F4, FC3, FCz, FC4, C3, Cz, C4, P3, P4, PO3, PO4, O1, and O2. In addition, horizontal electrooculogram (HEO) and vertical electrooculogram (VEO) electrodes were placed on the outer sides and above/below the eyes to monitor and correct for eye movement artifacts. During the experiment, EEG data were recorded at a sampling rate of 1000~Hz to ensure high temporal resolution. Electrode impedance was strictly maintained below 10~k$\Omega$ before recording to ensure signal quality and data reliability. Participants were instructed to remain as still as possible during the recording, minimizing body movement and eye blinks to reduce motion artifacts. Furthermore, a trigger electrode was used to synchronize stimulus event timing, and a computer monitor was employed to display experimental stimuli. Participants held a small numeric keypad to record their responses throughout the experiment. This equipment setup ensured the accuracy of data collection and the smooth execution of the experiment.

\subsubsection{Formal experimental process}

Before the experiment began, participants were seated, and the staff fitted them with the EEG cap and applied conductive gel to reduce impedance between the electrodes and the scalp. The staff then explained the key points of the experiment, including how to use the numeric keypad and maintain proper posture during the process.

Before the formal experiment, participants underwent a keyboard practice session to help them familiarize themselves with the experimental operations and to verify that all instruments were functioning properly. During the training phase, participants were shown images unrelated to the formal experiment on the screen. Then, text appeared on the screen instructing the participants on how to operate the keyboard, such as "Please press the 1 key after reading this message." Several trials of the same procedure as the formal experiment were also conducted to help participants become familiar with the specific details of the experimental process. This training phase used images unrelated to the selected experimental samples.

\begin{figure}[htbp]
    \centering
    \includegraphics[width=0.7\textwidth]{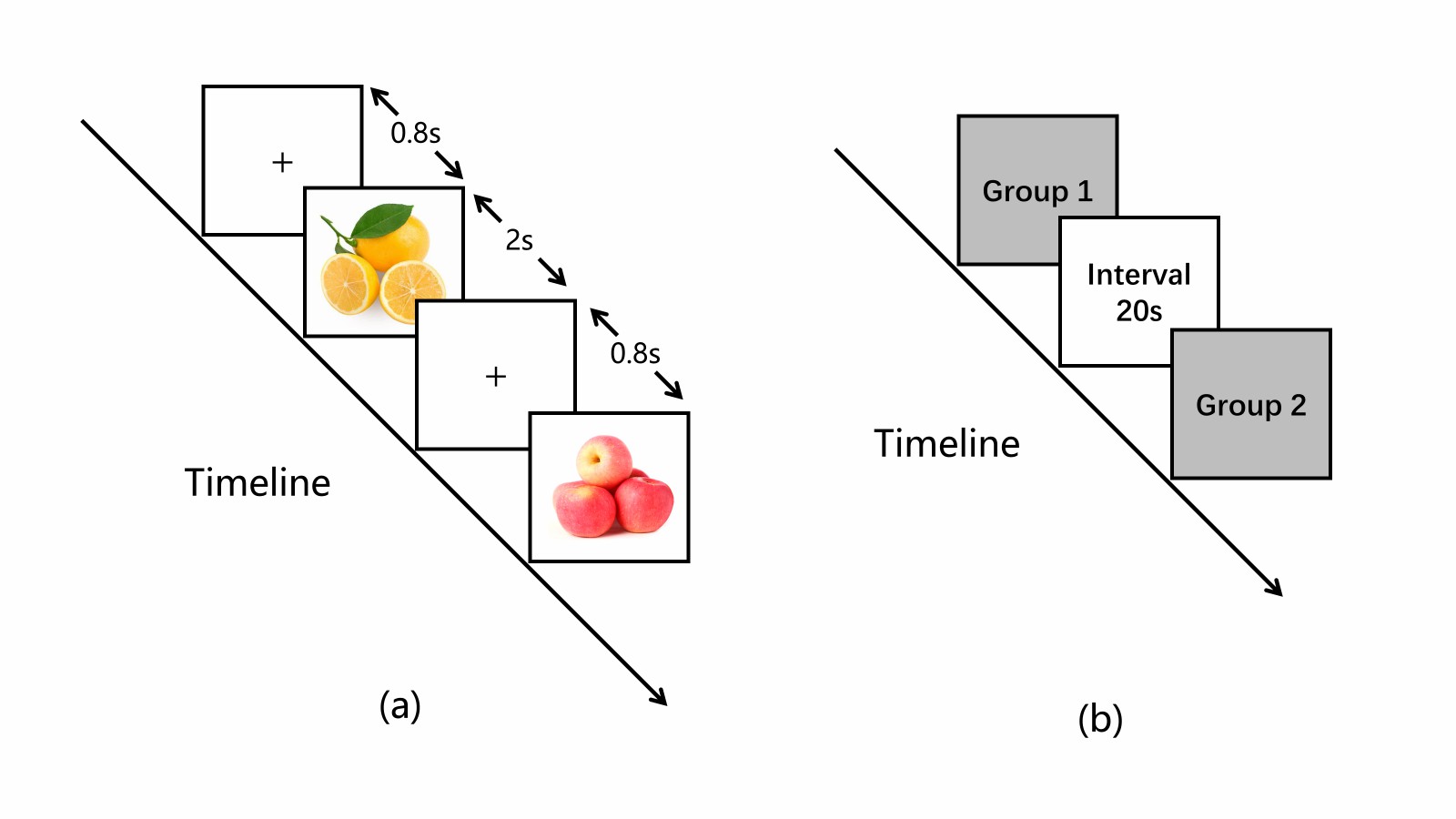}
    \caption{Experimental procedure. (a) Shows the time course within a single block. (b) Illustrates the relationship between blocks.}
    \label{fig:process}
\end{figure}

In the formal experiment, participants observed stimulus images on the screen and pressed the corresponding number buttons on the numeric keypad to indicate their preference for each image. The overall experimental procedure is shown in Figure \ref{fig:process}. The formal experiment was divided into six blocks, each presenting stimulus images from a specific category. There was a 20-second break between blocks to allow participants to rest and regain focus. Each block followed a specific sequence: first, a black cross was displayed in the center of a white background on the screen for 0.8 seconds. Then, a randomly selected image from the category was displayed for 2 seconds. Finally, a rating interface appeared, during which the participant used the keypad to rate the image. After the rating was completed, the interface disappeared. This cycle was repeated, moving on to the next loop with the same three steps. Once all the experiments were completed, a concluding screen informed the participants that the experiment had ended.

\subsubsection{Data Processing}

All data analysis was performed using MATLAB (R2018a, The MathWorks, Inc.) and the EEGLAB13 toolbox \cite{delorme2004eeglab}. First, since we focused primarily on specific brain regions and the absolute amplitude of potential changes, we applied offline mastoid re-referencing using the M1 and M2 electrodes to minimize potential artifact interference. Next, a band-pass filter with a frequency range of 1-30 Hz was applied to the data, and the data was segmented into time windows starting from 100 milliseconds before the stimulus onset to 2000 milliseconds after the stimulus onset. To further reduce artifacts, Independent Component Analysis (ICA) was performed using the FastICA algorithm \cite{pontifex2017variability, stone2002independent}. Based on the ICA results, eye blinks and other non-neural artifacts were identified and removed \cite{jung2000removal}. During data processing, baseline correction was applied using the 100-millisecond pre-stimulus period to correct for overall potential shifts and improve data accuracy.

In addition, averaging techniques were used to process the EEG data in order to exclude any non-specific voltage changes unrelated to the stimuli. Unrelated stimuli may trigger either positive or negative brain responses, which can be mutually canceled out over multiple measurements by averaging the results. This method effectively isolates neural activity directly related to the stimuli of interest, allowing us to focus on analyzing specific brain responses associated with the target stimuli. 

\section{Results}

\subsection{Behavioral Data Analysis}

The behavioral data recorded participants' ability to assess preferences during the pre-experiment survey, as well as their preferences for each specific sample during the formal experiment (based on their active selection via the keyboard).

\subsubsection{Survey Results}

Upon analyzing the survey data, we found significant differences in participants' ability to recognize and evaluate their product preferences. Specifically, only a small portion of participants (11.63\%) were able to clearly and consistently identify their preferences, meaning they could distinctly differentiate between products they liked and disliked. In contrast, more than half of the participants (51.16\%) reported that they could only distinguish their preferences to a moderate degree, showing a medium level of confidence and judgment ability. The remaining participants expressed difficulty in completing this task.

These results indicate that most people's emotional responses to products are not simply binary (like or dislike) but exist on a continuum of emotional states. Based on the survey results, we defined emotional states that neither explicitly express liking nor disliking as "neutral emotions." This refined classification method (like, dislike, neutral), derived from the survey results, not only enhances our understanding of the complexity of human emotions but also provides methodological support for further analysis of subtle differences in emotional responses.

\subsubsection{Formal Experiment}

\begin{figure}[htbp]
  \centering
  \includegraphics[width=0.8\textwidth]{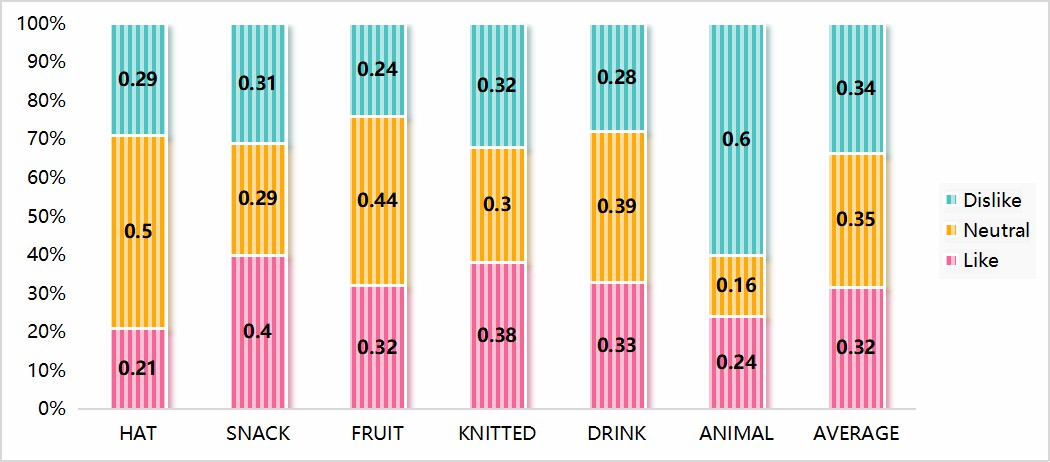}
  \caption{The preference ratio in six group}
  \label{fig:preference ratio}
\end{figure}

In the formal experiment, we conducted a statistical analysis of participants' preferences for different categories of images, as shown in Figure \ref{fig:preference ratio}. This result illustrates the distribution of preferences (like, neutral, dislike) across all six categories (hats, snacks, fruits, knitted items, beverages, and animals).

For the category of knitted items, the highest proportion of participants expressed liking (0.38), which may reflect the comfort or positive emotions related to warmth that knitted patterns evoke. In the case of animal images, the neutral response was the lowest (0.16), indicating that participants had more defined emotional preferences for animal images. This may be because animal images are often closely tied to personal experiences and emotional memories, leading participants to express clear preferences, either liking or disliking them. In contrast, hat images had the highest neutral response (0.5), suggesting that many participants held a neutral attitude toward hat patterns. This may be due to the generally neutral and widespread design of hats, which lack strong personalization and are less likely to evoke emotional associations, resulting in weaker emotional reactions. From the perspective of "dislike", in addition to the notable high proportion for animal images, snack images also had a relatively high dislike ratio (0.31), possibly influenced by concerns over food safety or personal dietary preferences.

Overall, the average data shows that the ratios for like, neutral, and dislike were 0.31, 0.35, and 0.34, respectively, indicating that participants' emotional responses to different categories of images were relatively balanced, with no single category significantly driving emotional tendencies across all participants. However, the differences in emotional responses across categories suggest that people's preferences and reactions to different types of visual stimuli are influenced by different neural systems and brain regions. This implies that preferences for different object categories may be affected by distinct brain areas \cite{10.3389/fpsyg.2019.00014}, likely including regions involved in emotion, memory, and aesthetic judgment.

\subsection{ERP analysis}

\subsubsection{Frequency Spectrum and Time-Domain Analysis}

\begin{figure}[htbp]
  \centering
  \includegraphics[width=0.9\textwidth]{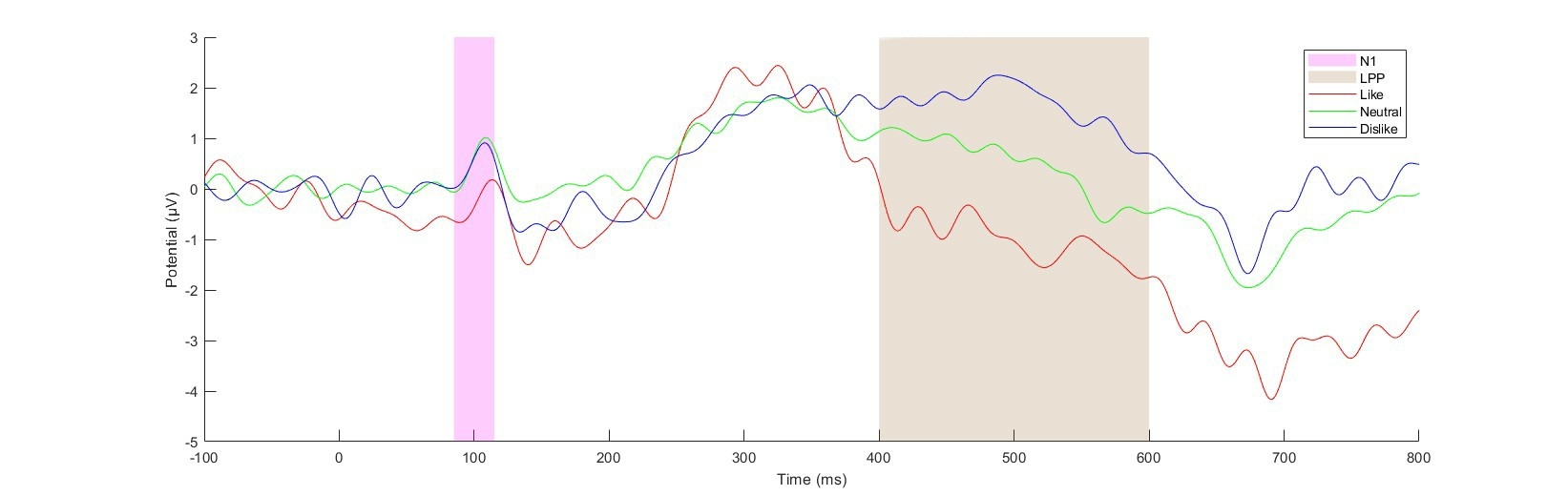}
  \caption{Grand average EEG waveforms under different preference conditions}
  \label{fig:EEG waveform}
\end{figure}

To minimize the influence of product types on preference analysis results, we employed a grand average method by aggregating EEG signals across all categories during both the spectrum and time-domain analyses. This approach averages EEG signals triggered by the same preference but different product types, effectively eliminating variations specific to product categories and allowing us to focus on EEG features associated with preferences themselves. Additionally, we averaged EEG signals across all channels for each participant. Figure \ref{fig:EEG waveform} shows the grand average EEG waveform across all participants, product types, and channels under different preference conditions (like, neutral, dislike), with N1 (purple-shaded area) and LPP (brown-shaded area) components marked in their respective time windows.

In Figure \ref{fig:EEG waveform}, the red line represents the average ERP waveform under the "like" condition, the green line represents the "neutral" condition, and the blue line represents the "dislike" condition. Since the N1 and LPP components are typically analyzed in specific channels, but this figure focuses on overall brain activity differences under different reference methods, we concentrate only on the positive and negative trends within the corresponding time windows for the grand average. During the N1 time window, the differences among the conditions (like, neutral, dislike) were minimal, suggesting that early sensory processing across the brain was fairly consistent. In contrast, during the LPP time window, the "dislike" condition showed a significant positive potential, while the "like" condition exhibited a pronounced negative potential. The "neutral" condition showed no significant difference compared to the mastoid-referenced EEG signals. We hypothesize that positive emotions, such as liking, tend to activate brain regions associated with reward and pleasure, particularly the prefrontal cortex, and manifest as negative potentials during the LPP time window. This explains why, under the grand average reference, the potential appears negative compared to the mastoid reference. On the other hand, stimuli associated with dislike often activate brain regions related to threat or aversion, such as the right parietal cortex, leading to positive potentials during the LPP window, thus shifting the grand average EEG toward positive values.

To comprehensively understand brain activity under different emotional states, we employed both frequency spectrum and time-domain analyses. These methods allowed us to analyze data across multiple sampling channels, rather than focusing solely on specific ERP components. Specifically, we selected the Fz and Cz channels for analysis, as these regions are associated with decision-making and cognitive attention, respectively \cite{cavanagh2014frontal}.

Through the Fz channel, we accessed EEG data from the prefrontal cortex, particularly the medial prefrontal region, which plays a key role in regulating and processing emotions \cite{ochsner2005cognitive, lewis2014functional}. Numerous studies have shown that neural activity in the prefrontal cortex is closely related to emotional experiences and management. Specifically, the Fz region exhibits clear electrophysiological responses when faced with positive or negative emotional stimuli. This region is also crucial for decision-making tasks, including risk assessment and moral judgment, with medial prefrontal activation being vital for completing such tasks. By analyzing the EEG signals from the Fz channel, we can delve deeper into how emotions influence decision-making processes, enhancing our understanding of this complex interaction.

In the Cz channel analysis, we focused on the central region of the brain, a critical area for processing perception and attention regulation \cite{pfurtscheller2010motor, taylor2015functional}. As part of the somatosensory cortex, the Cz region integrates sensory information from the entire body. Emotional changes can affect how this information is processed, which in turn influences individuals' perceptions and responses to their environment. Additionally, the Cz region plays a decisive role in attention allocation, with emotional fluctuations significantly impacting this process \cite{dolcos2010brain, anderson2005affective}. A detailed analysis of the Cz channel provides insights into how emotions regulate attention mechanisms and affect individuals' focus and processing efficiency during tasks. This analysis helps us further understand how emotional states modulate cognitive processes, especially in situations that require high levels of concentration and precise responses.

\begin{figure}[htbp]
    \begin{minipage}[t]{0.5\linewidth}
        \centering
        \includegraphics[width=\textwidth]{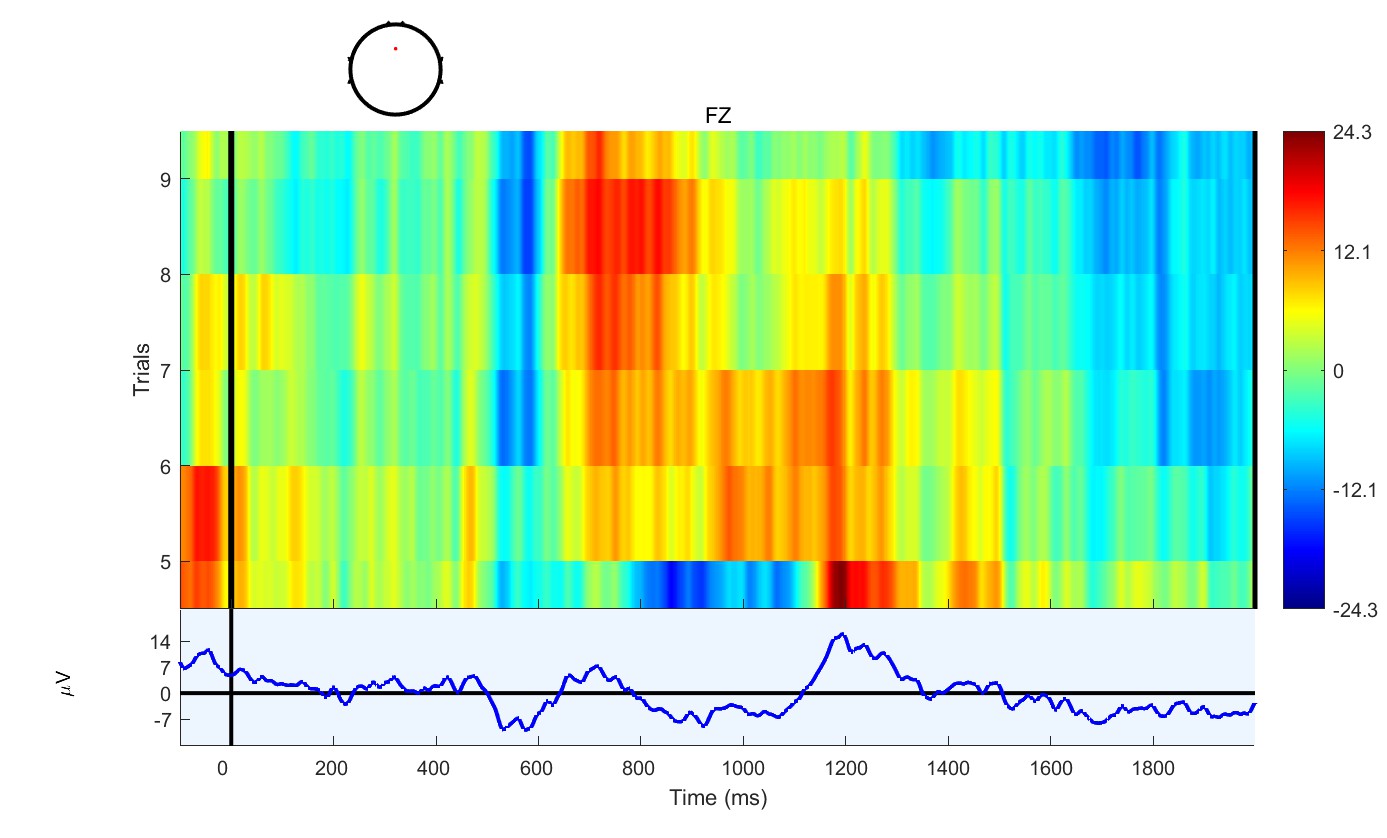}
    \end{minipage}%
    \begin{minipage}[t]{0.5\linewidth}
        \centering
        \includegraphics[width=\textwidth]{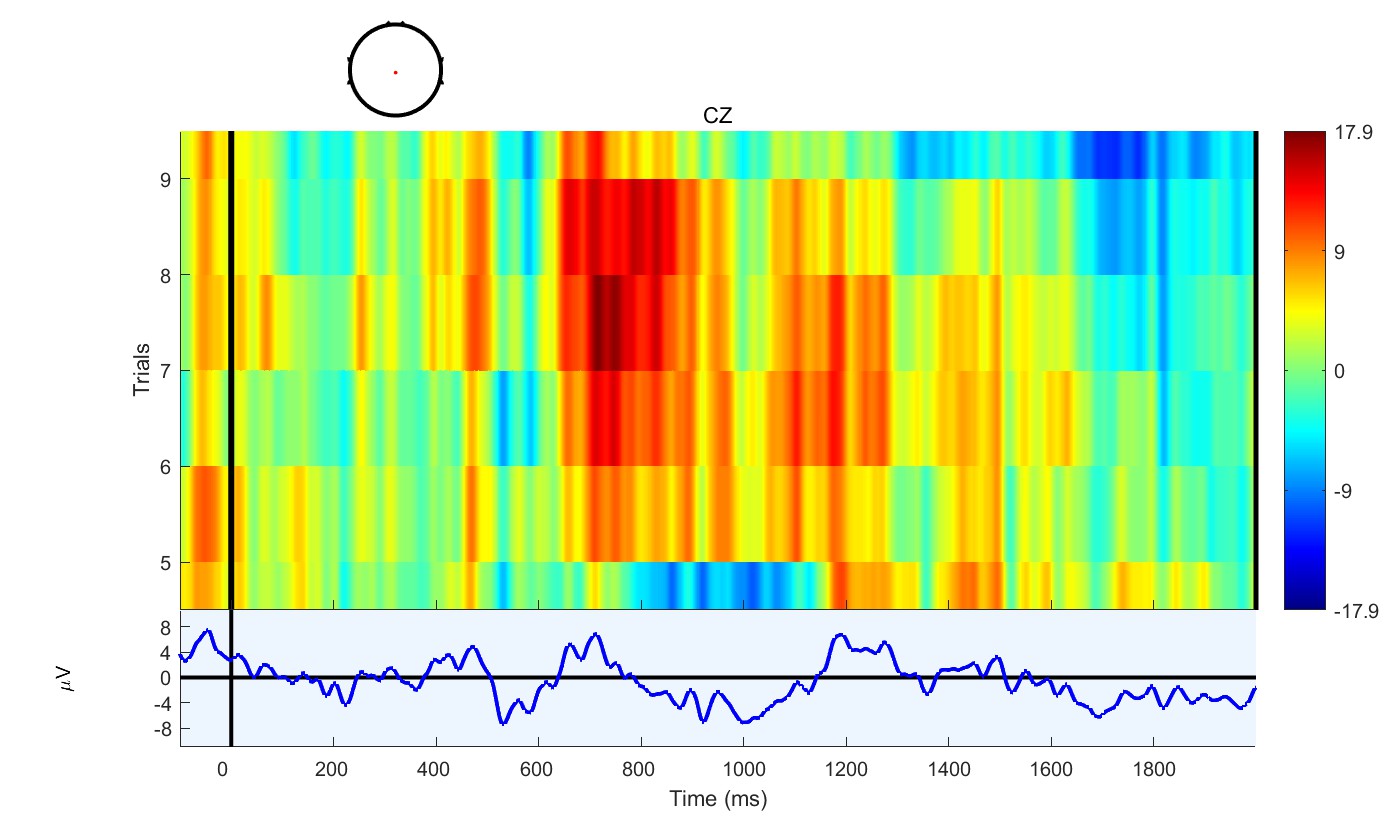}
    \end{minipage}
    \caption{ERP waveforms from Fz and Cz channels}
    \label{fig:FzCz}
\end{figure}

The left side of Figure \ref{fig:FzCz} shows the ERP image from the Fz channel. The heatmap at the top displays EEG activity across nine trials, illustrating voltage changes over time for each trial. The color bar ranges from -24.3~$\mu$V (blue) to 24.3~$\mu$V (red), representing different voltage levels. Around 0 ms, when the event starts, the heatmap shows a significant positive potential (red region), followed by negative potentials (blue regions) around 1000 ms and 1500 ms. These fluctuations may reflect electrophysiological responses during emotional regulation and processing. The right side of Figure \ref{fig:FzCz} shows the ERP image from the Cz channel, with the color bar ranging from -17.9~$\mu$V (blue) to 17.9~$\mu$V (red), indicating different voltage levels. Around 0 ms, a significant positive potential (red region) appears, followed by negative potentials (blue regions) around 800 ms and 1500 ms. These changes may reflect the electrophysiological responses of the perception and attention systems to emotional states. The waveform at the bottom of each figure shows the average voltage changes (ERP waveforms) across all trials. At 0 ms, a significant positive peak is observed, followed by a gradual decline in potential, with negative peaks around 800 ms and 1500 ms. These voltage changes highlight the central region's key role in perception and attention regulation. Similarly, the waveform at the bottom of the Fz channel shows a positive peak at 0 ms, followed by a gradual decline and a negative peak around 1000 ms. These voltage changes further confirm the importance of the prefrontal cortex in emotion and decision-making processes.

\subsubsection{N1 and LPP Component Analysis}

\begin{figure}[htbp]
  \centering
  \includegraphics[width=0.5\textwidth]{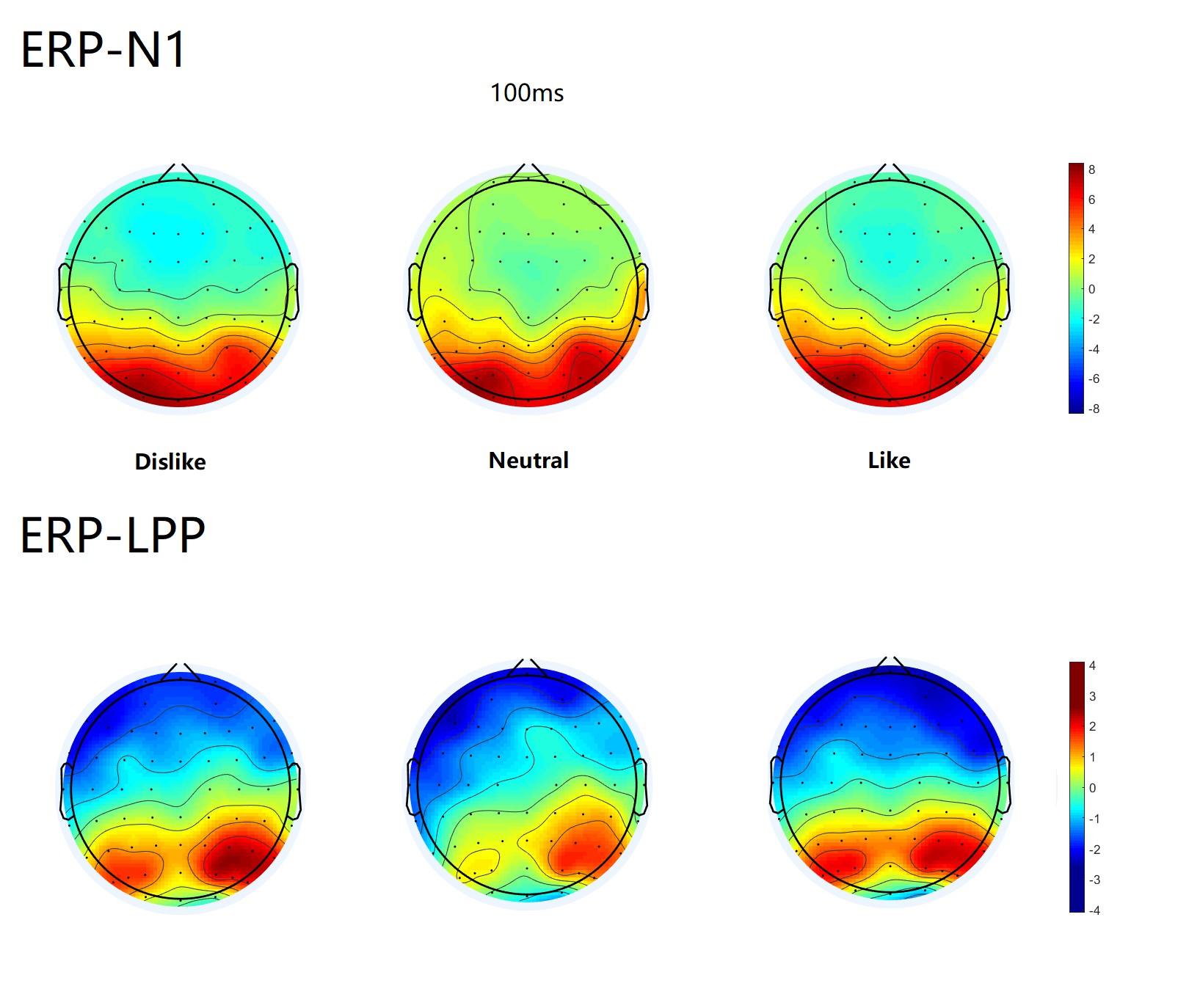}
  \caption{Scalp topography of ERP components at 100 ms and 500 ms under different preference conditions}
  \label{fig:ERP}
\end{figure}

N1 and LPP are two important components of Event-Related Potentials (ERPs). The N1 component typically appears within 90-110 milliseconds after stimulus onset, reflecting early visual attention processes \cite{mangun2013attentional}. The LPP component usually appears between 400-600 milliseconds, associated with emotional information processing and late cognitive evaluation \cite{schupp2012emotion, foti2010affective}. We hypothesized that there is a specific relationship between emotional preferences and these two ERP components. To analyze ERP components from the perspective of preference consistency, we performed a grand average across all EEG categories. Figure \ref{fig:ERP} shows the average voltage amplitudes corresponding to different preferences at 100 milliseconds and 500 milliseconds after the stimulus. These two time points correspond to the key activity windows of N1 and LPP, respectively. In these time windows, we conducted detailed analyses of the ERPs. Specifically, the N1 component was captured within the 90-110 millisecond window, primarily distinguishing between "like" and "neutral" emotional responses \cite{Foti2009DifferentiatingNR}, while the LPP component was captured within the 400-600 millisecond window, primarily identifying differences between "neutral" and other emotional types \cite{Hajcak2010}. These time windows were chosen based on previous studies, which found that the N1 component is related to visual attention and the LPP component is associated with emotional information processing.

Further analysis revealed significant differences in the N1 and LPP components in the frontal and central regions under different emotional states. Observing the N1 part in the upper section of Figure \ref{fig:ERP}, we can see that for the dislike condition (left panel), the negative potential was concentrated in the central region, with a relatively small overall amplitude. This suggests that during a "dislike" emotional state, participants exhibited lower early visual attention allocation, reflecting reduced sensitivity to emotionally evocative stimuli, which indicates a lower prioritization of processing for negative emotional stimuli. Additionally, the concentration of negative potentials was relatively localized, suggesting that the "dislike" emotion may not trigger widespread neural responses, supporting the attention-reduction effect often associated with negative emotions. In the neutral emotional state (middle panel), the distribution of negative potentials was more even, with an amplitude between those of the "like" and "dislike" emotions. While the distribution of negative potentials in the neutral state did not show significant concentration in specific regions, the even distribution indicates that the brain's processing of neutral emotional stimuli was relatively mild and widespread. This is consistent with the low emotional arousal of neutral stimuli, leading to non-preferential attention allocation across brain regions during early processing. For the like condition (right panel), the scalp map showed the most pronounced negative potentials in the frontal and central regions, particularly in the prefrontal area. This indicates that in the "like" emotional state, participants allocated more early visual attention, responding more strongly to emotional stimuli. The "like" state is typically associated with higher positivity and attention levels, and the strong negative potential in this state likely reflects the brain's prioritization of these stimuli. Therefore, the N1 component plays a crucial role in early emotional processing and serves as an important indicator of emotion-related attention mechanisms.

In the lower section of Figure \ref{fig:ERP}, which illustrates the LPP component, we observe that for the dislike condition (left panel), the LPP amplitude was relatively low, mainly concentrated in the posterior scalp region, and showed negative potential characteristics. This indicates that while "dislike" emotional stimuli could trigger some emotional responses, the late-stage emotional evaluation process was relatively weak, as the brain maintained lower sustained attention toward these stimuli. This phenomenon also reflects an important characteristic of the LPP component—its close relationship with emotional arousal. Stimuli with lower arousal levels tend to evoke weaker LPP responses. In the neutral condition (middle panel), the distribution of LPP was more uniform, with a slight increase in positive potential amplitude, but it remained at a relatively low level. This suggests that neutral emotional stimuli did not elicit significant emotional responses during late-stage emotional evaluation. While similar to the "dislike" condition in processing, the slightly higher positive potential indicates that the brain attempts to maintain cognitive balance and emotional stability when processing neutral stimuli. In the like condition (right panel), the positive potential of the LPP was most prominent in the frontal and central regions, with the largest amplitude in the prefrontal area. This indicates that the "like" emotional state not only had a significant effect during early visual attention processing (i.e., the N1 component), but also elicited strong responses during late-stage emotional evaluation. This phenomenon aligns with the characteristic of the LPP component: stimuli with high emotional arousal, whether positive or negative, tend to elicit stronger LPP responses, reflecting the brain's complex cognitive evaluation and emotional response at this stage.

\begin{figure}[htbp]
  \centering
  \includegraphics[width=0.9\textwidth]{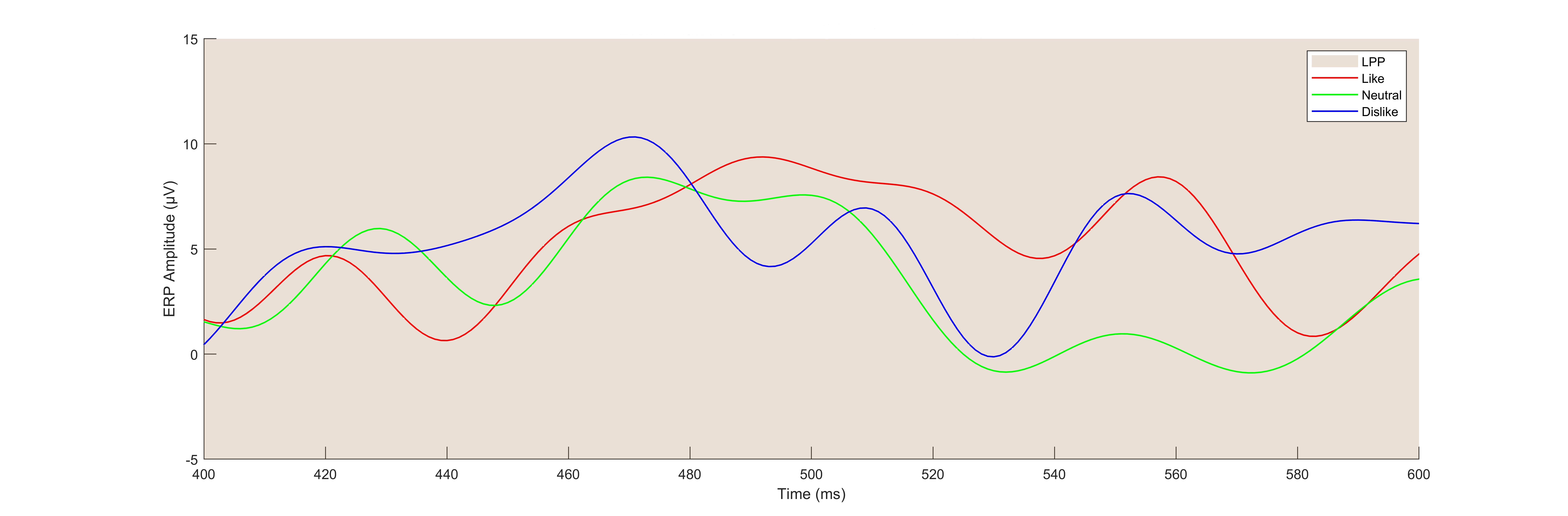}
  \caption{LPP potential changes under different preference conditions}
  \label{fig:LPP}
\end{figure} 

The LPP component is most prominent at electrode sites in the central and posterior brain regions. We selected central electrodes (C3, Cz, C4) to further investigate the relationship between LPP and emotional preferences. Figure \ref{fig:LPP} presents the EEG fluctuations in early and mid-LPP phases for different emotional preferences (like, neutral, dislike). The red line represents "like," the green line represents "neutral," and the blue line represents "dislike." The analysis shows that the neutral emotion had the smallest amplitude in the early LPP phase (before 500 milliseconds), and it returned to near-zero potential in the mid-LPP phase (after 500 milliseconds), with the smallest amplitude, indicating limited emotional processing of the stimuli. The "dislike" and "like" conditions were overall similar, though "dislike" had a slightly higher amplitude in the early LPP phase and a slightly lower amplitude in the mid-LPP phase than "like," indicating that the LPP component exhibits varying sensitivity to different emotional preferences.

The N1 component in the frontal and central regions was able to clearly distinguish between neutral and other emotions. Additionally, research indicates that in emotion-inducing stimuli, the N1 peak is highest for "like," followed by "dislike" and "neutral" emotions. This demonstrates the critical role of the N1 component in reflecting early stages of emotional processing, where greater interest or attention toward an image correlates with an increase in N1 potential amplitude. For LPP, stimuli with higher emotional arousal (whether positive or negative) elicit the highest peaks, indicating that the LPP plays an important role in the later stages of emotional response. When participants encountered neutral stimuli, the LPP waveform amplitude was relatively low, helping to distinguish emotional responses more accurately. In the frontal region, emotional differences in the LPP waveform were particularly pronounced, which may be related to the role of the prefrontal cortex in emotional processing. In the central-parietal region of the LPP waveform, although the differences between "like" and "dislike" emotions were not significant, there were still variations across different time periods.

Using both the N1 component from the early-to-mid stage and the LPP component from the mid-to-late stage of the overall cognitive process provides higher precision and reliability compared to using a single component or time window alone. Based on these two components, we can effectively distinguish between "like," "neutral," and "dislike" emotions. Studies have shown that the early visual attention mechanism reflected by the N1 component can quickly capture differences in emotional stimuli, while the late emotional processing mechanism represented by the LPP component reflects a more complex cognitive evaluation process. This dual-component analytical approach enables us to gain a more comprehensive understanding of the temporal dynamics and cognitive mechanisms underlying emotional responses.

\section{Discussion}

Identifying the relevant brain regions and ERP components is a key step in understanding the neural basis of human preference responses. Previous research has primarily focused on later ERP components, but it is important to note that early components like N1 can also provide valuable insights \cite{Nazari2014}. Several studies have demonstrated that N1 and LPP components can serve as indicators of preferences for various types of products, such as smartphones, humidifiers, and robots \cite{Guo_Ding_Wang_Liu_Jin_2016, Guo_Li_Chen_Duffy_2022, Guo2018AffectivePM}. In our study, we focused on N1 and LPP as key triggers that describe preference-related emotions. By analyzing the ERP waves, we observed that as electrode placement shifted towards the frontal coronal plane, the voltage amplitude decreased, indicating a spatial gradient of neural activity within the frontal cortex. Additionally, we found significant differences in whole-brain potentials during the LPP time window for different preferences. These findings highlight the potential effectiveness of N1 and LPP components as markers of preference-related ERP signals, providing insights into the neural processes underlying preference evaluation.

\subsection{ERP Component Selection}

In the existing literature, multiple ERP components have been shown to be associated with user preferences, particularly N1, P2, LPP, and N200, which have received widespread attention in various studies \cite{Guo_Li_Chen_Duffy_2022, Langeslag_van_Strien_2018, Wang2021HowDR}. Each of these components has distinct characteristics in visual tasks, plays a role in emotional processing, and exhibits sensitivity to emotional stimuli, making them important candidates for studying preference-related emotions. However, given the specific context of our experiment, after a series of comparisons and selections, we chose N1 and LPP as the primary components for analysis due to their high sensitivity to emotional preference changes and their strong discriminatory power.

First, the N1 component is considered to be closely related to early visual attention allocation, especially when processing task-relevant stimuli. N1 exhibits significant intensity dependence \cite{Luck1994, Salisbury2020}. N1 typically appears within 100 milliseconds after stimulus onset and reflects the brain's rapid response to perceptual information. Its negative potential shift indicates the degree of attention focus and the activation level of the perceptual system. For preference-related emotional research, the N1 component is particularly important because it can reflect the initial attention allocation when individuals are presented with stimuli they like or dislike. Our experimental results support this: when participants viewed preferred products, the N1 amplitude increased significantly, especially in the prefrontal region. This suggests that N1 is not only a marker of perceptual attention but can also serve as an early neural signal for emotional preferences. In contrast, neutral products elicited smaller N1 amplitudes, indicating that these stimuli did not attract significant attentional resources. It is worth noting that while N1 can serve as a criterion for preference judgment, it also appears in tasks with high responses and high task intensity, indicating its involvement in attention mechanisms \cite{Luck2000}.

The P2 component also attracted our attention. P2 usually appears within the 200-millisecond time window, and previous studies have shown that it is related to visual processing and emotional responses \cite{Feng2012, Guo_Li_Chen_Duffy_2022}. Specifically, P2 shows higher amplitudes for positive stimuli (such as attractive objects) and lower amplitudes for negative stimuli \cite{Wang2021HowDR}. However, in our experiment, the P2 component did not show enough discriminatory power. This might be because our experimental design focused on comparing choices within the same product category (e.g., durian versus lemon in the fruit category), rather than highlighting strong visual design or emotional stimuli. In this context, the sensitivity of the P2 component was less pronounced than N1, and thus it was not selected as a primary ERP marker.

Regarding N200, previous studies have found that it is often associated with error monitoring, conflict processing, and negative feedback \cite{Baker2011, Goto_Lim_Shee_Hatano_Khong_Buratto_Watabe_Schaefer_2019}. In particular, N200 amplitudes are typically larger when participants are exposed to disliked product stimuli \cite{Wang2021HowDR}. N200 responses are also closely linked to task-related processes, especially when participants have some expectation of upcoming stimuli \cite{Folstein2008}. However, in our experiment, the participants had no expectation or cues about the product stimuli, and the experiment was more focused on intuitive responses rather than task-driven reactions. As a result, the N200 component did not exhibit significant responses related to emotional preferences, and thus it was not selected as a major ERP component for preference analysis.

Finally, we chose the LPP component as another key component. LPP (Late Positive Potential) usually appears between 400-600 milliseconds and is widely recognized as a core indicator of emotional processing \cite{Hajcak2010}. LPP shows significant amplitude changes in response to emotionally arousing stimuli (both positive and negative), especially in processes related to attention allocation and emotional regulation \cite{Ferrari2011}. Our experimental results demonstrated that LPP showed significant amplitude differences in response to liked and disliked product stimuli, particularly in the central brain regions. Overall, disliked products tended to elicit smaller LPP responses, while liked products elicited larger positive potentials. This phenomenon can be explained by the fact that when individuals are faced with preferred stimuli, the brain allocates more cognitive resources and engages in more emotional evaluation, leading to increased LPP amplitudes.
Interestingly, in some cases, we observed that neutral stimuli triggered a larger mid-LPP response than disliked stimuli. This could be due to the ambiguity of human preferences, where neutral stimuli may require more complex cognitive evaluations to determine whether they are liked or disliked, thus leading to an increase in LPP amplitude \cite{Guo_Li_Chen_Duffy_2022}. This finding differs from previous studies on emotional images, which typically categorize stimuli as pleasant or unpleasant. In our experiment, neutral products required more emotional processing and cognitive judgment. This also suggests that relying solely on LPP analysis, which involves conscious processing, could lead to bias. Therefore, we adopted a combined analysis approach using both LPP and N1 components for a more comprehensive assessment.

\subsection{Neural Correlates of Preference Analysis}

Human preferences for different products are not simply perceptual reactions but involve multiple brain regions and complex neural mechanisms. By analyzing ERP signals from the prefrontal, central, and occipital regions, this study reveals the neural basis of preference-related emotions in the brain. Notably, the prefrontal cortex and parietal regions exhibited significant differential responses when processing preference-related emotions, findings that align with existing neuroscience literature.

The prefrontal cortex, particularly the right prefrontal area, is believed to play a central role in emotional regulation and decision-making \cite{Davidson1990}. In our experiment, we found that the N1 component in the prefrontal region showed a significant negative potential when participants viewed products they liked, while disliked products elicited a smaller N1 response. This finding can be interpreted as the prefrontal cortex allocating more cognitive resources for early visual processing and emotional evaluation when individuals encounter products they prefer. Research has shown that the prefrontal cortex is not only crucial in forming preference-related emotions but also plays a dominant role in resolving emotional conflicts and making complex decisions \cite{Sutton2010}. Thus, the activation of the N1 component in the prefrontal region can be seen as an early neural marker of preference-related emotions.

The LPP component in the central brain region further confirms the depth of this emotional processing. LPP is closely related to the intensity of emotional arousal, and its amplitude increased significantly when participants viewed both liked and disliked products, especially in the central and occipital regions. This indicates that in these regions, the brain not only processes emotional information but also engages in deep cognitive evaluation and emotional regulation. Studies have shown that the central and occipital brain regions are critical for visual perception and emotional information processing, particularly when dealing with complex emotional stimuli \cite{Ferrari2011}.

Additionally, we found that the LPP component was particularly pronounced in the right prefrontal and central regions. Under conditions of negative emotions, the right prefrontal cortex exhibited a stronger LPP response, suggesting that the right prefrontal cortex may play a more active regulatory role when processing negative emotions. This finding aligns with previous literature indicating the involvement of the right hemisphere in processing negative emotions \cite{Levy2011}. Future research could explore the distinct functional roles of the left and right hemispheres in processing emotional preferences, particularly how the cooperation between the hemispheres influences emotional responses when individuals face complex or conflicting emotions.
\subsection{Quantitative Model}

Based on the experiments described above and taking into account the common neural features of preference potentials triggered by different categories of stimuli, this study proposes a formula that reveals the quantitative relationship for evaluating preference responses across various stimuli. The model combines N1 and LPP components from different brain regions. All measurements are referenced to mastoid potentials, with units in microvolts (\(\mu V\)).

We define two variables for numerical analysis: Preference Intensity (\(I_P\)) and Preference Tendency (\(P_P\)), as well as the Preference Response Score (\(S_P\)).

Preference Intensity (\(I_P\)) represents the magnitude of the neural response to the stimuli. We expect neutral stimuli to produce lower values, while both "like" and "dislike" stimuli should produce higher values, without differentiating between positive and negative preferences.

From our experiments, we observed that generalized processing of stimuli is reflected in both the N1 and LPP components. In the N1 component, the prefrontal negative potential shift is less pronounced for neutral stimuli compared to "like" and "dislike" stimuli, with the values closer to the baseline. In the LPP component, the central-parietal positive potential decays more quickly for neutral stimuli compared to "like" and "dislike" stimuli, resulting in smaller absolute values for neutral emotions during the mid-LPP time window. Based on this, we propose the following relationship formula:

\begin{equation}
I_P = -F_{100} \times P_{550}
\end{equation}

Where \(F_{100}\) represents the prefrontal potential at 100 ms post-stimulus, and \(P_{550}\) represents the parietal potential at 550 ms post-stimulus. Multiplication is used in this formula to emphasize that when either potential value approaches zero, the overall preference intensity decreases, whereas larger values indicate a stronger preference intensity.

Preference Tendency (\(P_P\)) represents the degree of liking or disliking a stimulus, where positive values indicate "like" and negative values indicate "dislike". Due to the influence of noise and ambiguity in emotional judgment, the distinction between neutral and "like"/"dislike" emotions can be fuzzy. Hence, this variable focuses only on stronger emotional tendencies, ignoring neutral emotions.

In our experiments, the average brain potential during the LPP window closely matched this conceptual framework, with "like" and "dislike" stimuli exhibiting clear separation in positive and negative potentials. Additionally, considering the sequence differences between "like" and "dislike" in the central-parietal LPP component during the early and mid-phases, we introduced a correction based on the peak difference between these two emotions. The proposed relationship formula is:

\begin{equation}
P_P = -A_{500} + k \times (P_{530} - P_{440})
\end{equation}

Where \(A_{500}\) represents the average whole-brain potential at 500 ms post-stimulus, and \(k\) is a scaling factor, set to 0.05, to balance the potential difference between two time points. \(P_{530}\) and \(P_{440}\) represent the parietal potentials at 530 ms and 440 ms post-stimulus, respectively. The combination of addition and subtraction in this formula captures the changes in potential over time, providing a more comprehensive representation of emotional tendency.

Using Preference Intensity (\(I_P\)) and Preference Tendency (\(P_P\)), we can generate a pair of values representing a participant's response to a stimulus. Alternatively, we can simplify the analysis by calculating a single Preference Response Score (\(S_P\)):

\begin{equation}
S_P = \sqrt{|I_P|} \times P_P
\end{equation}

The square root in this equation serves to moderate the large variations in preference intensity (\(I_P\)) produced by multiplication, aligning it with the magnitude of the additive preference tendency (\(P_P\)). This helps to maintain numerical stability and simplifies the analysis.

With this set of relationships, we can reduce the complexity of EEG data into a single quantitative indicator, the Preference Response Score (\(S_P\)), or into a pair of values, making it easier to quickly analyze participants' preferences. This model could also be used for further affective computing or boundary delineation through machine learning methods.

\subsection{Applications of ERP in Emotion Preference Recognition and Future Prospects}

By analyzing the combination of N1 and LPP components in ERPs, this study demonstrates the potential of EEG technology in emotion preference recognition, particularly in fields such as personalized recommendations and intelligent device interactions. N1 and LPP components effectively capture users' responses in different emotional states, providing theoretical support for emotion recognition technologies based on EEG signals.

With the rapid development of Brain-Computer Interface (BCI) technology, EEG's role in real-time emotion monitoring is becoming increasingly important. For example, BCI-based personalized recommendation systems can analyze users' EEG signals and calculate there preference rapidly with our quantitative model to accurately recommend products, enhancing user experience. By identifying emotional responses, such systems can dynamically adjust recommendation content, improving the accuracy of personalized recommendations. Future intelligent devices could monitor users' EEG signals in real time and dynamically adjust device functions or environmental settings to respond to users’ needs in a more personalized manner. For instance, smart home systems could automatically adjust lighting, temperature, or music based on users' emotional states, creating a living environment that aligns more closely with their emotional needs. By decoding EEG signals, devices could better understand and predict users' emotional states, enabling more humanized interactions. This emotion-driven interaction model would significantly enhance user experience and provide technical support for emotional computing and autonomous adaptability in intelligent devices.

Although this study has made progress in emotion preference recognition, there are several limitations. First, the participant pool mainly consisted of university students, representing a relatively homogeneous social and cultural background, which may limit the generalizability of the findings. Future research should involve a more diverse participant pool, including individuals from different age groups, genders, and cultural backgrounds. Additionally, the study did not categorize and analyze specific stimuli within each item category, lacking a specialized understanding of each category. The experiment used only six types of everyday items as emotional stimuli, which may have varying degrees of emotional arousal. For example, snacks and pets may elicit stronger emotional responses due to their high emotional appeal, whereas items like hats or scarves may have more neutral attributes, making it harder to provoke noticeable emotional reactions. Future research could explore the impact of different levels of emotional arousal on ERP components by incorporating more categories and more complex stimuli. Moreover, using dynamic videos, virtual reality (VR), or augmented reality (AR) technologies could make experimental stimuli more immersive and reflective of real-life situations, enhancing the intensity and authenticity of emotional responses. In addition, combining EEG analysis with eye-tracking and functional magnetic resonance imaging (fMRI) could provide more comprehensive neurophysiological data, enabling a deeper understanding of the spatiotemporal dynamics of emotional responses, thereby further improving the accuracy and depth of emotion preference research.

This study primarily focused on N1 and LPP components; however, future research could further explore the role of other ERP components such as P200, N200, and P300. Different ERP components may serve distinct functions at various stages of emotional response. By incorporating multiple ERP components, it would be possible to build a more comprehensive emotion recognition model. For example, the P200 component is typically associated with early emotional evaluation, while the N200 component may be linked to conflict monitoring and error detection \cite{Folstein2008}. The temporal relationships between these components in emotional processing can reveal the dynamic process of emotional responses, helping to decode the formation mechanisms of complex emotions more precisely.

Overall, ERP technology holds great promise in emotion preference recognition. By expanding participant diversity, enriching experimental stimuli, and integrating more ERP components, future research could further enhance the accuracy and real-time capability of emotion recognition, providing technical support for personalized services and emotional computing in intelligent devices. This not only advances the field of affective computing but also presents new opportunities for emotion-driven intelligent systems.

\section{Conclusion}

This study for the first time combines the analysis of both the N1 and LPP components of Event-Related Potentials (ERP) to explore the neural mechanisms underlying human preferences. A visual stimulus dataset encompassing multiple categories of everyday items was designed for the experiment. Through time-domain and frequency-domain analyses of EEG signals, it was found that the N1 component reflects early attention allocation in the prefrontal cortex during preference judgments, while the LPP component represents late cognitive processing in emotional evaluation in the central and posterior parietal regions. These components exhibited significant differences across different time windows and brain regions, particularly under conditions of "like," "dislike," and "neutral" emotional states, revealing the neural dynamics associated with preference-related emotions.

Based on the significance of the N1 and LPP components, this study proposes a quantitative relationship model integrating early visual processing and late emotional evaluation to compute preference intensity and tendency. The experimental results demonstrated that preferred stimuli elicited higher N1 negative potentials and LPP positive potentials, while disliked stimuli showed lower N1 amplitudes and pronounced LPP responses. The development of this model simplifies complex EEG data, providing a novel tool for emotion recognition technologies and uncovering the distinct roles of various brain regions in preference-related emotions.

In summary, this study reveals the dynamic neural response characteristics of the brain when processing preference-related emotions through a detailed combined analysis of the N1 and LPP components. By employing multi-category visual stimuli and conducting multi-region analyses, a quantitative model was proposed to offer new theoretical support for emotion recognition technologies. This research not only lays the groundwork for future studies on EEG-based preference classification but also establishes a robust foundation for advancing brain-computer interface technologies.

\section*{Acknowledgments}
This work was supported by the National Social Science Fund of China (No. 20BYY177).

\bibliographystyle{unsrt}
\bibliography{reference}

\end{document}